\documentclass[sigconf]{acmart}

\usepackage{amsmath,booktabs,array,graphicx,tikz,float,balance,colortbl}

\AtBeginDocument{%
  }

\setcopyright{acmlicensed}
\copyrightyear{2018}
\acmYear{2018}
\acmDOI{XXXXXXX.XXXXXXX}
\acmISBN{978-1-4503-XXXX-X/2018/06}

\definecolor{pendingred}{RGB}{190,20,30}

\newcommand{\ds}[1]{\textcolor{black}{D#1}}

\newcommand{\A}{\mathcal A}
\newcommand{\C}{\mathcal C}
\newcommand{\E}{\mathcal E}
\newcommand{\setenc}{\operatorname{SetEnc}}
\newcommand{\gru}{\operatorname{GRU}}

\begin{document}

\title{Learning Complementary Agent Groups via Task Hypergraphs for Next-POI Recommendation with Missing Modalities}

\author{Jinze Wang}
\email{jinzewang@swin.edu.au}
\affiliation{%
  \institution{Swinburne University of Technology}
  \city{Melbourne}
  \state{VIC}
  \country{Australia}
}

\author{Yuze Liu}
\affiliation{%
  \institution{Swinburne University of Technology}
  \city{Melbourne}
  \state{VIC}
  \country{Australia}}

\author{Liang Yuan}
\affiliation{%
  \institution{Australian Catholic University}
  \city{Sydney}
  \state{NSW}
  \country{Australia}
}

\author{Gaoyuan Du}
\affiliation{%
 \institution{University of Tennessee}
 \city{Knoxville}
 \state{Tennessee}
 \country{USA}}

 \author{Xiaohang Xu}
\affiliation{%
  \institution{University of Tokyo}
  \city{Tokyo}
  \country{Japan}}

\author{Yong Xiang}
\affiliation{%
  \institution{Deakin University}
  \city{Melbourne}
  \state{VIC}
  \country{Australia}}

\author{Jiong Jin}
\affiliation{%
  \institution{Swinburne University of Technology}
  \city{Melbourne}
  \state{VIC}
  \country{Australia}}



\renewcommand{\shortauthors}{Trovato et al.}

\begin{abstract}
Next point-of-interest (POI) recommendation uses visit histories and POI content, but text and images are often missing. Large language model agents have been applied to recommender systems for preference modeling and evidence acquisition. The challenge is to select agents that can jointly analyze the available evidence for each request. We propose \textsc{CATH} (\emph{Complementary Agents via Task Hypergraphs}), which learns to select agent groups for tasks such as comparing candidates or analyzing a POI with missing content. Agents with distinct processing roles are the task hypergraph nodes; each hyperedge connects the selected group and records its analysis goal and target POIs. The policy selects members conditioned on the request, task, and agents already selected. During execution, members share candidate-linked results that guide their subsequent operations. Completed tasks update candidate states and agent memories within the request. A shared ranking head converts candidate states into corrections to a fixed base ranking. Supervised ranking trains the executor, followed by policy optimization that accounts for computation cost. Experiments under natural missingness on \textcolor{black}{Gowalla-CA, Foursquare-NYC, and Foursquare-TKY show 8.4--9.1\% relative NDCG@20 gains over HyperAgent4POI, the strongest prior baseline evaluated}. Additional analyses compare group-selection and execution designs, test ranking under additional masking, and characterize quality--cost tradeoffs.

\end{abstract}




\maketitle

\section{Introduction}
Location-based Web services rank places using users' visit histories and heterogeneous point-of-interest (POI) content~\cite{zheng2024decentralized,chen2025next,wu2025decentralized,wang2026large,dong2026towards,wang2021footprint}. Conventional next-POI recommenders learn sequential and spatiotemporal patterns from check-in histories \citep{luo2021stan,chen2025enhancing,wang2025generative}. Multimodal methods add POI text and images to enrich these signals \citep{xu2024mmpoi,du2023cross,zhang2026mscl,dong2026language}. Such content can distinguish nearby candidates with similar categories, but is often incomplete. Missing content leaves a choice about how to analyze the remaining evidence: \textbf{which historical visits or POI attributes matter for the current request, and which analyses should work together?}

Large language models (LLMs) support semantic analysis of POI content and user histories~\cite{zhuang2026think2go,lv2026reasoning,wang2025we}. Recommendation research has explored LLM-based agents for preference modeling \citep{zhang2024agentcf,wu2025mas4poi}, evidence acquisition through tools \citep{wang2026meta,huang2025recommender,wang2024recmind,dong2025ha}, and request-conditioned POI representations \citep{wang2026agent4poi,cheng2025poi,wang2025beyond}. In our setting, an \emph{agent} is a decision unit that chooses operations and maintains state, using an LLM for semantic analysis. Its \emph{role} specifies its processing capabilities, accessible inputs, and permitted operation types.

Within one prediction request, a \emph{task} specifies an analysis goal and target candidates, such as summarizing preferences for the full pool, comparing two POIs, or examining a candidate with missing content. An \emph{agent group} contains agents selected for that task. Members may perform several operations; all tasks contribute to the same next-visit prediction.

\begin{figure}[t]
\centering
\includegraphics[width=\linewidth]{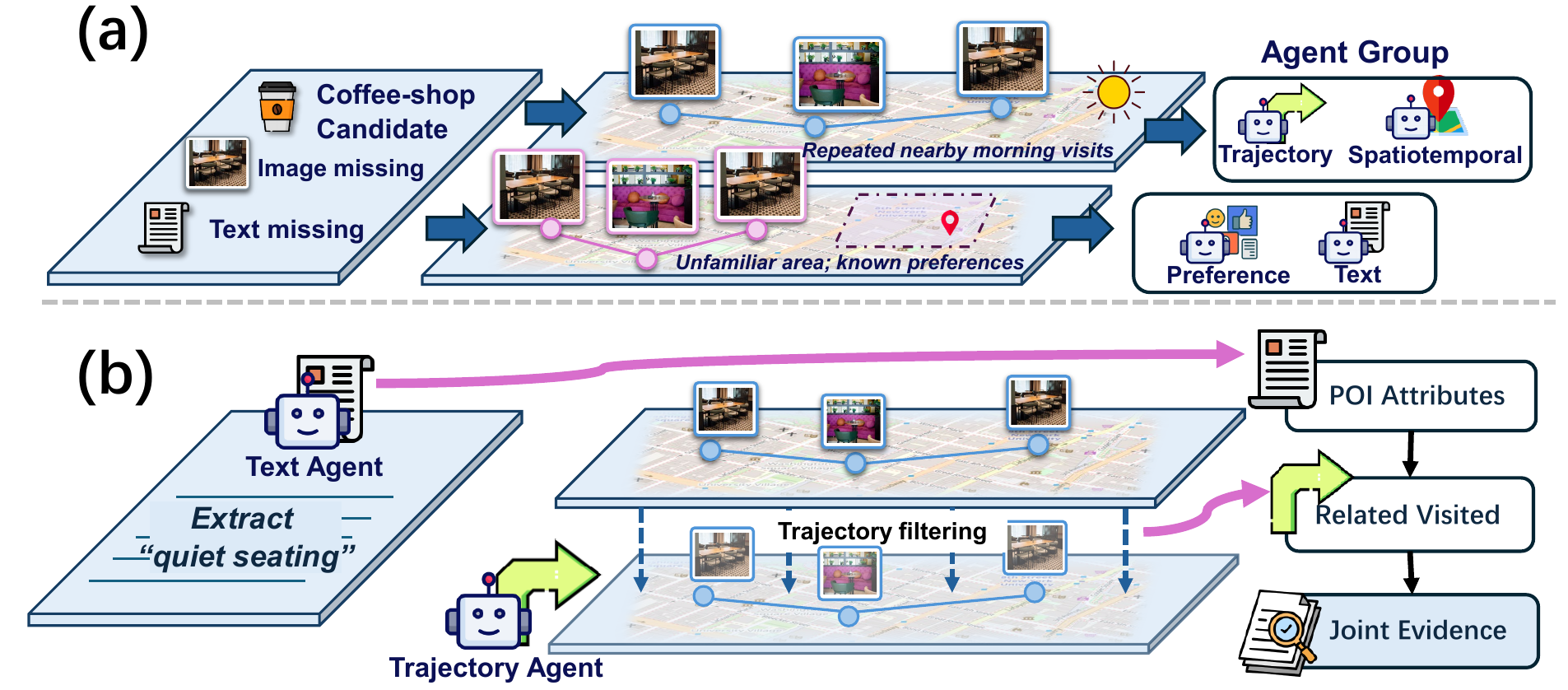}
\caption{Illustrative agent-group choice and shared evidence. (a) The same missing-image candidate can call for different agents under different histories. (b) A text attribute guides which past visits the trajectory agent compares.}
\label{fig:motivation}
\Description{Repeated nearby visits suggest trajectory and spatiotemporal agents, while unfamiliar-area requests suggest preference and text agents. Shared text evidence guides a later trajectory comparison.}
\end{figure}

Consider the same coffee-shop candidate with a missing image (Figure~\ref{fig:motivation}a). Repeated nearby morning visits make trajectory and spatiotemporal analysis useful for one user. For another user unfamiliar with the area, available text can reveal POI attributes, and preference analysis can relate them to past visits. A modality mask identifies available inputs, but history and task determine which processing capabilities are useful. Our history-stratified analysis illustrates this dependence: the ordering of agent groups changes across request strata under the same masking condition (Section~\ref{sec:ablation}).

Semantic messages also support missing-modality completion on recommendation hypergraphs \citep{wang2026hyperagent4poi}. Our focus is how to select agents for a common recommendation task and use their shared evidence to guide subsequent operations on its target candidates.

We aim to select agents whose operations combine useful evidence for the request. Two agents may return overlapping evidence, while another pairing may connect information that neither can assess alone. For example, a text agent can extract a ``quiet seating'' attribute, but its relevance depends on the user's behavior. Sharing that attribute lets a trajectory agent focus on related visits before comparing the candidate with alternatives (Figure~\ref{fig:motivation}b). We connect member selection to a shared task state, which records collected evidence and guides subsequent operations.

We propose \textsc{CATH} (\emph{Complementary Agents via Task Hypergraphs}), which represents agent groups with \emph{task hypergraphs} (Figure~\ref{fig:overview}). Ordinary graph edges connect pairs of nodes; a hyperedge can connect several selected members in one relation. Each request has a task hypergraph whose nodes are agents. Each hyperedge represents a task instance with its selected agent group, goal, and target POIs. The POIs are the objects being analyzed. A learned decision rule, the \emph{policy}, first chooses a task and targets, then grows the group through invitations conditioned on the existing members. Once membership is fixed, execution begins. Members share candidate-linked results that guide their next retrieval or comparison. Local task states separate these exchanges, while participating agents carry request-level memory into later tasks. A shared scoring function, or ranking head, converts the accumulated evidence into a correction to a fixed base ranking.

The \emph{executor}, which performs operations and updates states, first learns from observed next visits. With this executor fixed, policy optimization learns task, member, operation, and stopping decisions from ranking utility and computation cost. Budgets bound the work spent on each request. Compared with clustering-derived agent groups \citep{liu2025hygma} or single-controller tool acquisition \citep{wang2026meta}, our core novelty is this task-oriented recommendation interface: candidate targets and goals govern member selection, candidate-aligned shared evidence guides subsequent operations, and joint results directly update a common candidate ranking head.

Our contributions are:
\begin{itemize}
\item We introduce CATH, which formulates multimodal next-POI reranking as task-level agent-group selection. Each task links an analysis goal and target candidates to members selected under the available evidence and budget.
\item We connect candidate-linked shared results to members' subsequent operations and a common ranking function. Task-local states organize joint execution, while agent memories carry evidence between tasks within the same request.
\item We evaluate the system on three benchmarks under natural and artificial missingness, compare alternative selection and execution designs, and analyze ranking quality against measured computation cost.
\end{itemize}

\section{Related Work}
\paragraph{Next-POI recommendation and incomplete content.}
Flashback retrieves recurrent hidden states using spatiotemporal similarity \citep{yang2020location}; STAN attends to spatial and temporal relations \citep{luo2021stan}; GETNext combines a trajectory flow map with a transformer \citep{yang2022getnext}. MMPOI integrates multimodal semantics and check-in sequences through a trajectory graph and adaptive multi-task transformer \citep{xu2024mmpoi}. LRMM uses modality dropout and an autoencoder for rating prediction \citep{wang2018lrmm}. HyperAgent4POI refines soft incidence and missing-modality representations through semantic messages on user/POI hypergraphs, then caches representations for online ranking \citep{wang2026hyperagent4poi}. Our task hyperedges represent task instances and their selected agents within an individual reranking request; their shared state guides concrete retrieval and comparison operations.

\paragraph{Request-conditioned processing and agent interaction.}
Agent4POI constructs request-conditioned multimodal POI representations at recommendation time \citep{wang2026agent4poi}. Meta-Modal Agent (MMA) formulates reranking as budgeted sequential evidence acquisition: tool returns enter the observation history used for subsequent routing and stopping \citep{wang2026meta}. AgentCF models users and items as language agents that jointly reflect on interaction records and update their preference memories \citep{zhang2024agentcf}. Our agents have processing roles and maintain states for a common analysis goal and target candidates. Their candidate-linked shared results enter each member's next operation. Participating agents retain evidence across overlapping tasks within one request, and their memories reset between requests.

\paragraph{Experts and agent groups.}
HMoRec combines hypergraph construction with mixtures of experts \citep{zhou2026enhanced}, and MAGNET uses interaction\allowbreak-conditioned routing with modality-role experts \citep{dai2026modality}. DyLAN performs team optimization followed by dynamic task solving \citep{liu2023dynamic}. HYGMA groups agents by history similarity and supplies group-aware features to action policies \citep{liu2025hygma}. In contrast to cluster-derived groups or single-agent sequential routing \citep{wang2026meta}, our architecture establishes a concrete recommendation interface: groups are formed dynamically for target candidates and subtask goals, candidate-aligned shared results guide members' subsequent operations within the task hyperedge, and joint task outcomes directly update a common residual ranking head while retaining episodic memory (Sections~\ref{sec:selection}--\ref{sec:ranking}). Set Transformer \citep{lee2019set}, proximal policy optimization (PPO) \citep{schulman2017proximal}, and learned stopping \citep{bacon2017option} supply the set encoder and policy components for this formulation.

\section{Problem Formulation}
\label{sec:problem}
One recommendation request asks which POI a user will visit next, using information available at prediction time $t_0$. We begin with a retrieved pool and improve its ordering through additional analysis. Formally, the request and candidate pool are
\begin{equation}
q=(u,\mathcal H_u^{<t_0},\xi_q),\qquad \C_q=\{p_1,\ldots,p_K\},
\end{equation}
where $u$ is the user, $\mathcal H_u^{<t_0}$ the observed visit history, and $\xi_q$ the location and time available at prediction. The unknown next-visit time is excluded. A fixed base sequence recommender supplies $\C_q$ and initial scores $r_p^{\mathrm{base}}$. Training pools are generated without fitting to the current request's next-visit label. The pool remains unchanged during reranking.

The work for one request forms an episode. An \emph{agent} is a decision unit that selects operations and maintains state. Its fixed \emph{role} specifies processing capabilities, permitted inputs, and operations. Within the episode, a \emph{task} specifies an analysis goal and the candidates to examine. An \emph{agent group} contains the agents selected for a task. An \emph{operation} is a concrete action within the analysis, such as retrieving a history segment or comparing candidates. A task may require several operations and exchanges among its members. The index $t$ counts task steps within the request, distinct from visit time.

POI content comprises text, images, categories, and geographic attributes. The mask $M_{p,m}=1$ denotes an available modality; $M_p$ collects these indicators for POI $p$. Candidate and historical POIs retain their masks, distinguishing unavailable content from observed zero-valued features. Frozen content encoders, a sequence encoder, and temporal and spatial features supply the initial request and candidate representations. Selection reads these representations, masks, operation availability, and accumulated analysis states. Detailed analysis and targeted retrieval occur after agents are selected.

Evaluation is chronological. Histories, indices, and timestamped content follow the request cutoff. Content without timestamps enters under an explicit static-side-information assumption. Artificial masking precedes caching and indexing. At inference, the policy, target generator, and tools have no access to the held-out next-visit label. The test pool is the unmodified retriever output, so the true next POI $p^+$ appears only if retrieved. Reranking can improve next-visit ranking metrics only for these covered requests. Retrieval coverage, conditional reranking, and end-to-end ranking consequently measure distinct aspects of performance (Section~\ref{sec:evaluation}).

\section{CATH: Agent Groups on Task Hypergraphs}
\label{sec:method}
\begin{figure*}[t]
\centering
\includegraphics[width=\linewidth]{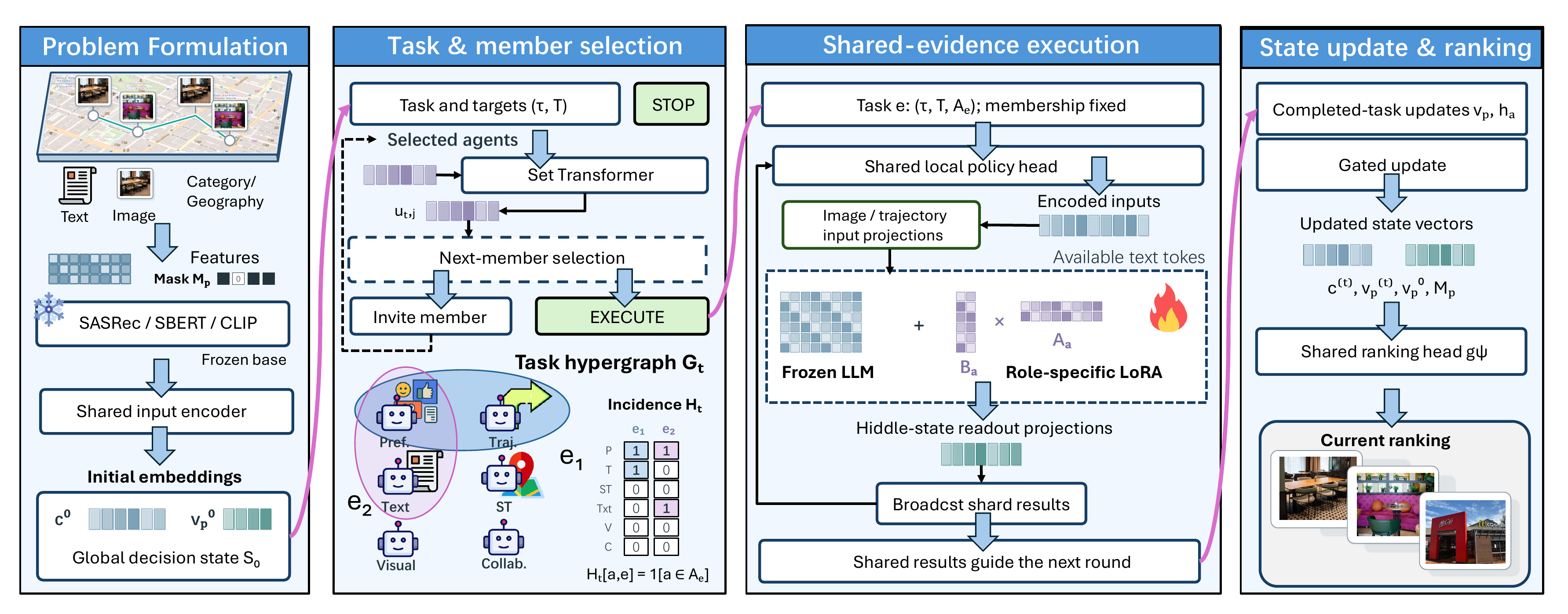}
\caption{CATH workflow from available inputs to POI ranking. The policy selects tasks and targets, then forms agent groups represented as task hyperedges. With membership fixed, agents execute operations through a shared frozen LLM and role-specific adapters. Shared results guide subsequent operations. Completed tasks update agent memories and candidate states, which a shared ranking head uses to recompute scores.}
\label{fig:overview}
\Description{A flowchart shows input encoding, task and member selection, shared-evidence execution, and state updates for ranking. Illustrative hyperedges e1 and e2 share the preference agent, with membership recorded by an incidence matrix. A shared local policy chooses operations for fixed members, and a frozen LLM with role-specific adapters produces their outputs. Broadcast results guide subsequent rounds. Completed tasks update agent memories and candidate states for the shared ranking head.}
\end{figure*}

For each request, CATH's policy chooses an analysis task, forms an agent group, and runs that group for one or more rounds. Members share candidate-related results between rounds and use them to choose subsequent operations. Completed tasks update the ranking and agent memories before the policy chooses another task or stops. Figure~\ref{fig:overview} summarizes input encoding, task and member selection, shared-evidence execution, and state updates for ranking.
\subsection{What a task asks the agent group to analyze}
A task is a pair $(\tau,T)$: the task type $\tau$ specifies the analysis goal, and the target set $T\subseteq\C_q$ specifies the affected candidates. We use three types (Table~\ref{tab:tasks}). A request-state update analyzes user context for the full pool. A candidate comparison examines the relative suitability of a legal pair. Semantic compensation examines one candidate whose content is incomplete, using the evidence that remains available. These types separate request-wide context, local ranking ambiguities, and candidate-specific content gaps so the policy can choose the scope of computation.

For example, a request-state update can summarize recurring activities for all candidates. A comparison can examine two adjacent coffee shops. Compensation can examine an image-missing shop using available text and relevant history. Members can perform several operations to answer the task's question. Compensation refines a ranking representation from available evidence; the unavailable raw modality and its mask remain unchanged.

Only executable task--target pairs are offered. Comparison pairs come from adjacent candidates in the current ranking, optionally supplemented by adjacent within-category pairs, with duplicates removed. They are regenerated after reranking. Compensation targets incomplete-content candidates individually, including low-ranked candidates. All routing controls share these rules.
\begin{table}[!t]
\centering\small
\setlength{\tabcolsep}{2.5pt}
\caption{Task types and target scopes. A task can contain several role-specific operations and exchange rounds.}
\label{tab:tasks}
\begin{tabular}{@{}>{\raggedright\arraybackslash}p{0.30\linewidth}>{\raggedright\arraybackslash}p{0.15\linewidth}>{\raggedright\arraybackslash}p{0.49\linewidth}@{}}
\toprule
Task type $\tau$ & Targets $T$ & Analysis goal \\
\midrule
Request-state update & Full pool $\C_q$ & Analyze user context for all candidates.\\
Candidate comparison & Legal pair & Compare the relative suitability of two candidates.\\
Semantic compensation & One POI & Analyze an incomplete-content candidate using available evidence.\\
\bottomrule
\end{tabular}
\end{table}

\subsection{Agents, task hypergraphs, and states}
The set $\A$ contains six agent nodes, one for each fixed processing role: long-term preference, short-term trajectory, spatiotemporal analysis, text analysis, visual analysis, and collaborative retrieval. Each agent can participate in multiple tasks. Its role determines permitted inputs and operations. Each agent maintains request-level memory and task-local state and uses a shared policy head to choose its operations. A shared frozen LLM backbone, role-specific low-rank adapters (LoRA), and trainable projections implement the agents' analysis interfaces.

The preference agent uses pre-prediction history and available content to match recurring interests with candidates. The trajectory agent examines recent visit order, categories, and legal time gaps, and can retrieve similar segments. The spatiotemporal agent uses known location and time, distances, and historical transitions. The text agent accesses available descriptions and reviews; the visual agent accesses available candidate or historical images. The collaborative retrieval agent searches permitted training histories and POI indices for similar users, trajectories, or related POIs. Thus, the visual agent can remain eligible for an image-missing target when relevant historical images exist, while the target image stays unavailable. Task targets, data permissions, and budget restrict each agent's legal operations.

Selecting an agent group $A_e\subseteq\A$ for $(\tau,T)$ creates one task instance $e$. The same group may serve other task instances. We record each instance as a hyperedge connecting its agent members:
\begin{equation}
\mathcal G_t=(\A,\E_t,H_t),\qquad H_t[a,e]=\mathbf1[a\in A_e].
\end{equation}
Here $\E_t$ contains the hyperedges representing task instances formed so far within the request, and $H_t$ is the agent--hyperedge incidence matrix. Each hyperedge stores its task type and targets. A hyperedge can contain any legal nonempty agent group, including a singleton. Membership stays fixed during its execution; later hyperedges can overlap. The incidence relation determines whose messages enter an analysis, who can read its detailed shared results, and whose memory is updated when it finishes.

In Figure~\ref{fig:overview}, hyperedge $e_1$ connects preference and trajectory agents, while $e_2$ connects preference and text agents. The preference agent is a shared node, but the two columns of $H_t$ record different memberships. Each hyperedge retains its own goal, targets, and task state. The task hypergraph therefore records joint analyses and their overlapping participants within one request.

The stored states follow these scopes. The request state $c^{(t)}$ summarizes user context, while $v_p^{(t)}$ stores the current representation of candidate $p$. A shared input encoder initializes $c^{(0)}$ and $v_p^{(0)}$ from the representations in Section~\ref{sec:problem}. Each agent has request-level memory $h_a^{(t)}$ of its completed work. During task $e$, it also has task-local state $l_{a,e}$ for intermediate work. The agent group maintains a task summary $\bar z_e$ and candidate-specific results $z_{e,p}$. Separate local states let a reused agent work on a new task without overwriting another task's intermediate results. Agent memories reset between requests. The global decision state is
\begin{equation}
S_t=\Bigl(c^{(t)},\{v_p^{(t)},M_p,r_p^{(t)}\}_{p\in\C_q},
\{h_a^{(t)}\}_{a\in\A},\{\bar z_e\}_{e\in\E_t},B_t\Bigr),
\end{equation}
where $r_p^{(t)}$ is the current score and $B_t$ the remaining budget. This state supplies the task and agent selectors. Within a task, each agent observes its permitted inputs, task and targets, shared results, and own state. An agent cannot read another agent's unsent private results.

\subsection{Selecting tasks and their members}
\label{sec:selection}
At task step $t$, the policy selects STOP or a legal pair $(\tau,T)$. For a selected task, it chooses an initiator and adds members sequentially. The set $A_{t,j}$ contains the members selected after invitation step $j$. A Set Transformer produces a permutation-invariant summary of that set:
\begin{equation}
u_{t,j}=\setenc\!\left(\{[\operatorname{role}(a),h_a^{(t)},
\operatorname{avail}(a,\tau,T)]:a\in A_{t,j}\}\right).
\end{equation}
Here $\operatorname{role}(a)$ encodes agent $a$'s fixed role, and $\operatorname{avail}(a,\tau,T)$ describes that agent's availability for the task and targets. The next-member distribution is
\begin{equation}
\pi_\theta(a\mid S_t,\tau,T,A_{t,j})=
\operatorname{softmax}_{a\in\mathcal V_{t,j}}
 f_\theta(S_t,\tau,T,u_{t,j},h_a^{(t)},\operatorname{role}(a)),
\label{eq:invite}
\end{equation}
where $f_\theta$ scores the legal choices $\mathcal V_{t,j}$. They contain unselected eligible agents and EXECUTE. The latter uses a dedicated action embedding and starts the current nonempty agent group. Invited agents join deterministically. Conditioning on $u_{t,j}$ lets the next choice account for the selected agents' capabilities, as well as the request and task. Selection uses known capabilities and accumulated states before the unselected agents execute. Training retains the actual invitation sequence although the set summary is permutation invariant. Once EXECUTE is chosen, the resulting set becomes $A_e$ and remains fixed throughout the task.

\subsection{Shared evidence guides subsequent operations}
\label{sec:execution}
A task initializes its shared and local states from its targets, members, and available request state. The index $r$ counts execution rounds within this fixed-membership task. Members choose operations from the same round-start snapshot. The system executes them, aggregates their outputs, and broadcasts the updated result before the next round. Feedback conditions the members' subsequent operation choices; member selection resumes only after the task ends. Candidate pairs, history ranges, and other tool parameters come from deterministic legal choices; empty retrievals return an explicit status.

Consider the illustrative comparison of two coffee shops by preference and text agents. The text agent can extract an available attribute, such as quiet seating, while the preference agent examines relevant visits. Neither agent sees the other's unsent output in that round. After broadcast, the preference agent can use the attribute to select more relevant history in its next operation. The candidates and members remain fixed, while the operations respond to the shared result.

Each operation returns a candidate vector $m_{a,e,p}$, a summary, coverage, status, and measured cost. These outputs use a learned vector interface. Available text enters the executor as tokens, while encoded images and trajectories enter through continuous projections. Hidden states at candidate readout positions are projected into $m_{a,e,p}$; a separate readout produces the member's task summary. The vectors carry evidence for aggregation, and the common ranking head produces the scalar scores. Coverage identifies which candidates were actually processed, so a partial result contributes only to its covered candidates.

Let $\mathcal D_{e,p}^{(r)}$ contain the latest valid messages covering $p$ from selected members, with role, operation, and status metadata. A permutation-invariant message encoder $\setenc_E$ summarizes them, and a gated recurrent unit (GRU) updates the candidate result:
\begin{equation}
z_{e,p}^{(r+1)}=
\begin{cases}
\gru_E\bigl(z_{e,p}^{(r)},\setenc_E(\mathcal D_{e,p}^{(r)})\bigr),
 &\mathcal D_{e,p}^{(r)}\ne\varnothing,\\
z_{e,p}^{(r)},&\text{otherwise}.
\end{cases}
\label{eq:edge}
\end{equation}
An uncovered candidate retains its previous result. A new agent--target message replaces its previous message, rather than creating an additional member contribution. The task summary combines candidate results and member summaries. After broadcast, the readout $\operatorname{Read}$ combines agent memory with the shared summary and candidate results for the local-state update:
\begin{equation}
l_{a,e}^{(r+1)}=\gru_A\!\left(l_{a,e}^{(r)},
\operatorname{Read}\bigl(h_a^{(t)},\bar z_e^{(r+1)},\{z_{e,p}^{(r+1)}\}_{p\in T_e}\bigr)\right).
\end{equation}
A shared local policy head, separate from the LLM executor, maps the updated member state to its next operation type:
\begin{equation}
P(o_{a}^{(r+1)} \mid l_{a,e}^{(r+1)}) = \frac{\exp\bigl(w_{o}^{\top} [l_{a,e}^{(r+1)}, \operatorname{role}(a)]\bigr)}{\sum_{o' \in \mathcal{O}_a} \exp\bigl(w_{o'}^{\top} [l_{a,e}^{(r+1)}, \operatorname{role}(a)]\bigr)},
\label{eq:local_policy}
\end{equation}
where $o_{a}^{(r+1)} \in \mathcal{O}_a$ is an operation type permitted by agent $a$'s role (e.g., attribute extraction, trajectory filtering, candidate comparison, or FINISH). All members share this head evaluated on their respective local states. Tool execution parameters (such as target candidates in $T_e$ and retrieval ranges) are determined deterministically from the task context without stochastic parameter sampling. An agent can also ask an active member to analyze specified candidates in a later round. The task ends when all members finish, no legal work remains, or its execution limit is reached.

When the task ends, gated updates write its summary into the participating members' memories. These memories carry completed analysis to later tasks, while detailed intermediate results remain scoped to their task. Candidate results update the targeted ranking states:
\begin{equation}
v_p^{(t+1)}=\gru_P\bigl(v_p^{(t)},[z_{e,p},\bar z_e,c^{(t)}]\bigr),\quad p\in T_e.
\end{equation}
Non-target states remain unchanged. Request-state-update tasks also update $c^{(t+1)}$ and cover the full pool. Compensation uses available evidence while retaining the original missingness masks.

Across multiple tasks, participating agents carry completed analysis in their episodic memories into subsequent tasks, while common candidate states preserve earlier ranking updates.

\subsection{Integrating task results into ranking}
\label{sec:ranking}
After each completed task, one shared ranking head reads the request and candidate states. The learnable function $g_\psi$ produces a scalar correction for each candidate:
\begin{align}
r_p^{(t)}={}&r_p^{\mathrm{base}}+
g_\psi(c^{(t)},v_p^{(t)},v_p^{(0)},M_p)\nonumber\\
&-g_\psi(c^{(0)},v_p^{(0)},v_p^{(0)},M_p).
\label{eq:rank}
\end{align}
The second term is evaluated at the current states; the final term is the initial-state reference. They cancel at initialization, leaving the base ranking unchanged. Later tasks alter the states, and the head recomputes the correction. All candidates share this scoring function, including candidates untouched by the current task. Their scores depend on the retained candidate state and current request context. Thus, task outputs enter a common ranking through state updates instead of independently calibrated agent scores.

The policy may stop immediately and return the base ranking. Legal-action masks reserve enough budget for an operation, and execution limits bound each task and request. Cost is normalized computation measured in a fixed environment, including routing, aggregation, retrieval, LLM forwards, and content re-encoding. Preprocessing is separated from per-request cost, cache hits incur their actual cost, and tokens, seconds, and calls are reported separately.

\newcommand{\pr}[2]{\textcolor{black}{\ensuremath{\text{#1}_{\pm#2}}}}
\begin{table*}[!t]
\centering\small
\caption{End-to-end ranking at the main missingness condition (natural absence). Cells report means over 3 seeds; subscripts give $\pm$ seed std. R/N denote Recall/NDCG; higher is better. \ds{1}=Gowalla-CA, \ds{2}=Foursquare-NYC, \ds{3}=Foursquare-TKY.}
\label{tab:main}
\setlength{\tabcolsep}{2pt}
\begin{tabular}{@{}>{\raggedright\arraybackslash}p{.25\linewidth}*{9}{>{\centering\arraybackslash}p{.073\linewidth}}@{}}
\toprule
& \multicolumn{3}{c}{\ds{1}} & \multicolumn{3}{c}{\ds{2}} & \multicolumn{3}{c}{\ds{3}}\\
\cmidrule(lr){2-4}\cmidrule(lr){5-7}\cmidrule(lr){8-10}
Method & R@20 & N@20 & MRR & R@20 & N@20 & MRR & R@20 & N@20 & MRR\\
\midrule
Frozen base
  & \pr{.318}{.004} & \pr{.192}{.003} & \pr{.148}{.004}
  & \pr{.271}{.005} & \pr{.164}{.004} & \pr{.128}{.003}
  & \pr{.284}{.004} & \pr{.174}{.003} & \pr{.136}{.003}\\
Flashback
  & \pr{.362}{.006} & \pr{.219}{.005} & \pr{.168}{.004}
  & \pr{.312}{.007} & \pr{.186}{.005} & \pr{.148}{.005}
  & \pr{.325}{.005} & \pr{.197}{.004} & \pr{.154}{.004}\\
STAN
  & \pr{.381}{.005} & \pr{.231}{.004} & \pr{.178}{.004}
  & \pr{.328}{.006} & \pr{.198}{.004} & \pr{.157}{.004}
  & \pr{.342}{.005} & \pr{.208}{.004} & \pr{.163}{.003}\\
GETNext
  & \pr{.412}{.005} & \pr{.252}{.004} & \pr{.194}{.004}
  & \pr{.354}{.006} & \pr{.215}{.004} & \pr{.170}{.004}
  & \pr{.371}{.005} & \pr{.227}{.003} & \pr{.178}{.003}\\
LRMM (adapt.)
  & \pr{.388}{.006} & \pr{.234}{.005} & \pr{.182}{.004}
  & \pr{.331}{.007} & \pr{.198}{.005} & \pr{.159}{.005}
  & \pr{.347}{.006} & \pr{.210}{.004} & \pr{.165}{.004}\\
HMoRec (adapt.)
  & \pr{.421}{.005} & \pr{.256}{.004} & \pr{.198}{.004}
  & \pr{.358}{.006} & \pr{.217}{.004} & \pr{.172}{.004}
  & \pr{.374}{.005} & \pr{.228}{.004} & \pr{.178}{.003}\\
MAGNET (adapt.)
  & \pr{.418}{.005} & \pr{.258}{.004} & \pr{.198}{.004}
  & \pr{.361}{.006} & \pr{.218}{.004} & \pr{.174}{.004}
  & \pr{.378}{.005} & \pr{.232}{.003} & \pr{.181}{.003}\\
HyperAgent4POI
  & \pr{.447}{.004} & \pr{.276}{.003} & \pr{.213}{.003}
  & \pr{.389}{.005} & \pr{.237}{.004} & \pr{.188}{.003}
  & \pr{.405}{.004} & \pr{.250}{.003} & \pr{.196}{.003}\\
MMA-Auto (adapt.)
  & \pr{.431}{.005} & \pr{.264}{.004} & \pr{.204}{.004}
  & \pr{.372}{.006} & \pr{.226}{.004} & \pr{.179}{.004}
  & \pr{.388}{.005} & \pr{.238}{.004} & \pr{.187}{.003}\\
Agent4POI (adapt.)
  & \pr{.438}{.005} & \pr{.271}{.004} & \pr{.209}{.004}
  & \pr{.380}{.006} & \pr{.232}{.004} & \pr{.183}{.004}
  & \pr{.396}{.005} & \pr{.244}{.003} & \pr{.192}{.003}\\
\midrule
Independent scoring
  & \pr{.456}{.004} & \pr{.282}{.003} & \pr{.218}{.003}
  & \pr{.396}{.005} & \pr{.242}{.003} & \pr{.192}{.003}
  & \pr{.412}{.004} & \pr{.255}{.003} & \pr{.200}{.003}\\
\rowcolor{black!6}
CATH
  & \pr{.484}{.004} & \pr{\textbf{.301}}{.003} & \pr{.232}{.003}
  & \pr{.419}{.005} & \pr{\textbf{.257}}{.003} & \pr{.204}{.003}
  & \pr{.437}{.004} & \pr{\textbf{.271}}{.003} & \pr{.213}{.003}\\
\bottomrule
\end{tabular}
\end{table*}

\section{Learning from Recommendation Feedback}
\label{sec:learning}
Training separates learning to produce useful evidence from learning to choose computation. Warm-up teaches the role adapters, projections, and ranking modules to translate executed tasks into candidate scores. Policy optimization then learns task, member, operation, and stopping decisions with these modules fixed, providing a stable scoring interface for comparing choices.

\paragraph{Utility and return.}
The positive item $p^+$ is the user's actual next visit recorded in the training data; it is unknown when making a prediction. For training requests whose positive item is retrieved, define the ranking utility
\begin{equation}
U_t=\log P_\psi(p^+\mid q,\C_q,S_t)
=r_{p^+}^{(t)}-\log\sum_{p\in\C_q}\exp r_p^{(t)}.
\label{eq:utility}
\end{equation}
A higher utility means more probability is assigned to the true next POI, even when its rank is unchanged. After a task completes, the reward is
\begin{equation}
R_t=U_{t+1}-U_t-\lambda_{\mathrm{cost}}c_t.
\label{eq:return}
\end{equation}
Here $c_t$ includes selection and execution cost. The coefficient $\lambda_{\mathrm{cost}}$ sets the penalty per unit of computation; the episode budget separately caps admissible work. A task earns reward through the change in the shared prediction, giving its members a common outcome rather than separate agent-specific accuracy targets. A request's return sums these rewards across completed tasks and stopping. The undiscounted return equals final utility improvement minus total cost. The next-visit label is read only by the supervised loss and reward evaluator after execution; decision policies, the return estimator, and tools receive prediction-time inputs. Evaluation and validation model selection use ranking metrics rather than this training utility.

\paragraph{Executor warm-up.}
Legal task trajectories supervise execution parameters $\phi$ (projections and role adapters) and aggregation/ranking parameters $\psi$ (message encoders, gated updates, and scoring head). They minimize $\mathcal L_{\mathrm{rank}}=-U$ at completed task states. Samples cover missingness patterns, singleton and multi-member groups, and no-new-content execution. With discrete actions fixed, gradients pass through execution and aggregation. The base recommender, content encoders, and LLM base weights remain frozen.

\paragraph{Selection warm-up from executed combinations.}
Executed legal agent combinations provide warm-up records under comparable states, tasks, targets, initiators, and local-operation rules. A set evaluator $Q_\omega$ is fitted to their measured net utility changes and ranks these recorded combinations for policy initialization. Its supervision comes only from combinations that have been executed. Higher-utility combinations supply invitation sequences with a fixed initiator and variable ordering of the other members. The evaluator is used during training; the learned policy selects groups at inference. Complete trajectories can also warm up task, operation, and stopping heads.

\paragraph{Discrete policy optimization.}
We freeze execution, ranking, and state-update parameters ($\phi,\psi$), then optimize the discrete policy with PPO \citep{schulman2017proximal}. Agent memories and candidate states still update as tasks execute. A legal-action mask removes unavailable choices under the permitted inputs, current membership, and remaining budget before action probabilities are normalized. A shared critic estimates the expected return from the current state using permitted global information. Advantage estimation compares observed returns with this estimate, connecting earlier invitation, operation, and stopping decisions to later task rewards. Policy encoders are trained separately from the frozen executor.

\section{Experiments}
\label{sec:evaluation}
We examine four questions: whether agent-group collaboration improves ranking with missing content, how request histories inform group selection, how shared-result execution compares with alternative organizations, and what online computation the resulting ranking requires.

\subsection{Experimental Setup}
\begin{table}[!t]
\centering\small
\caption{Benchmark scale, content availability, and fixed-pool ranking across three datasets. Content percentages precede artificial masking. Cond./E2E: covered-query/all-query NDCG@20.}
\label{tab:data}\label{tab:coverage}
\begin{tabular*}{\linewidth}{@{\extracolsep{\fill}}lrrr@{}}
\toprule
Statistic & \ds{1} & \ds{2} & \ds{3}\\
\midrule
Users & \textcolor{black}{6,442} & \textcolor{black}{1,083} & \textcolor{black}{2,293}\\
POIs & \textcolor{black}{13,889} & \textcolor{black}{38,333} & \textcolor{black}{61,858}\\
Visits & \textcolor{black}{171,623} & \textcolor{black}{227,428} & \textcolor{black}{573,703}\\
Test queries & \textcolor{black}{7,251} & \textcolor{black}{2,248} & \textcolor{black}{5,763}\\
Text (\%) & \textcolor{black}{61.3} & \textcolor{black}{72.1} & \textcolor{black}{68.4}\\
Image (\%) & \textcolor{black}{34.7} & \textcolor{black}{41.8} & \textcolor{black}{38.2}\\
\midrule
Coverage (\%) & \textcolor{black}{89.2} & \textcolor{black}{87.6} & \textcolor{black}{91.3}\\
Base (Cond./E2E) & \textcolor{black}{.215/.192} & \textcolor{black}{.187/.164} & \textcolor{black}{.191/.174}\\
Independent & \textcolor{black}{.316/.282} & \textcolor{black}{.276/.242} & \textcolor{black}{.279/.255}\\
CATH & \textcolor{black}{.337/.301} & \textcolor{black}{.293/.257} & \textcolor{black}{.297/.271}\\
\bottomrule
\end{tabular*}
\end{table}

\paragraph{Data and candidate protocol.}
Following \citep{yang2022getnext}, we evaluate on \textcolor{black}{Gowalla-CA, Foursquare-NYC, and Foursquare-TKY}, with dataset statistics reported in Table~\ref{tab:data}. We refer to these datasets as \ds{1}, \ds{2}, and \ds{3}, respectively, throughout the tables. Consistent with prior work on incorporating textual and visual POI content \citep{xu2024mmpoi,zhang2026dima,wang2025beyond}, we augment these check-in datasets with matched textual descriptions and static images. Each user's visits are divided chronologically into \textcolor{black}{80\%/10\%/10\%} training, validation, and test partitions. A frozen \textcolor{black}{SASRec} retriever supplies the same \textcolor{black}{100} candidate POIs to every method; targets outside this pool are not inserted. Histories and retrieval indices follow the prediction cutoff. A missingness mask records which content is available. Natural absence persists; artificial masks hide some observed text or images before caching and indexing.

\paragraph{Baselines and controls.}
Flashback, STAN, and GETNext model next-POI trajectories \citep{yang2020location,luo2021stan,yang2022getnext}. LRMM, HMoRec, MAGNET, and HyperAgent4POI provide incomplete-content comparisons \citep{wang2018lrmm,zhou2026enhanced,dai2026modality,wang2026hyperagent4poi}; MMA-Auto and Agent4POI provide learned evidence routing and request-dependent processing \citep{wang2026meta,wang2026agent4poi}. All adaptations score the common pool using legal next-POI inputs. MMA-Auto has no clarification channel, and Agent4POI receives no extra intent or social annotations. \textcolor{black}{We use author-released code or our adaptations.
Selection controls compare choices of members with a shared executor and reward. Execution controls compare organizations for the same task and targets; the exchange/no-exchange pair also fixes membership. Complete-request systems choose their own tasks and groups, so they can visit different states and measure the final behavior of the full system. Table~\ref{tab:ablation} reports parameter counts and realized costs.}

\paragraph{Metrics and implementation.}
For a rank cutoff $L$ and evaluated request set $\mathcal Q$, let $k_q$ be the true next-POI rank, with $k_q=\infty$ for a retrieval miss:
\begin{equation}
\begin{aligned}
\mathrm{Recall}@L
&=\frac{1}{|\mathcal Q|}\sum_q\mathbf1[k_q\le L],\\
\mathrm{NDCG}@L
&=\frac{1}{|\mathcal Q|}\sum_{q:k_q\le L}\frac{1}{\log_2(k_q+1)}.
\end{aligned}
\label{eq:evalmetrics}
\end{equation}
NDCG rewards ranking the next POI higher; mean reciprocal rank (MRR) averages $1/k_q$. All three assign zero to retrieval misses. \textcolor{black}{NDCG@20 is the primary metric}, weighting requests equally. Coverage is the fraction of requests whose next POI is retrieved; conditional scores average covered requests, while end-to-end scores average all requests (Table~\ref{tab:coverage}). Table~\ref{tab:main} reports means and standard deviations over \textcolor{black}{3 seeds ([42, 123, 456])}; differences use percentage points (pp).

\begin{table}[!t]
\centering\small
\setlength{\tabcolsep}{2pt}
\caption{Selection and execution of agent groups. Columns show NDCG@20 on \ds{1}--\ds{3} and NYC cost. (a) Joint 30\% masking; shared task/executor; selector parameters. (b) Natural absence; fixed tasks/targets, with group fixed for exchange/no-exchange; trainable parameters.}
\label{tab:ablation}\label{tab:exchange}
\begin{tabular*}{\linewidth}{@{\extracolsep{\fill}}lccccc@{}}
\toprule
Configuration & \ds{1} & \ds{2} & \ds{3} & Cost & Params\\
\midrule
\multicolumn{6}{@{}l}{\textit{(a) Group selection}}\\
Fixed group
  & \textcolor{black}{.264} & \textcolor{black}{.226} & \textcolor{black}{.238}
  & \textcolor{black}{0.48} & \textcolor{black}{2.1M}\\
Mask-only
  & \textcolor{black}{.271} & \textcolor{black}{.232} & \textcolor{black}{.245}
  & \textcolor{black}{0.51} & \textcolor{black}{2.4M}\\
Independent
  & \textcolor{black}{.279} & \textcolor{black}{.237} & \textcolor{black}{.251}
  & \textcolor{black}{0.54} & \textcolor{black}{3.8M}\\
Direct group
  & \textcolor{black}{.281} & \textcolor{black}{.240} & \textcolor{black}{.254}
  & \textcolor{black}{0.56} & \textcolor{black}{3.4M}\\
CATH (set-cond.)
  & \textcolor{black}{\textbf{.293}} & \textcolor{black}{\textbf{.250}} & \textcolor{black}{\textbf{.264}}
  & \textcolor{black}{0.61} & \textcolor{black}{4.2M}\\
\midrule
\multicolumn{6}{@{}l}{\textit{(b) Shared-result execution}}\\
No exchange + fusion
  & \textcolor{black}{.284} & \textcolor{black}{.241} & \textcolor{black}{.255}
  & \textcolor{black}{0.54} & \textcolor{black}{4.8M}\\
Single controller
  & \textcolor{black}{.292} & \textcolor{black}{.249} & \textcolor{black}{.263}
  & \textcolor{black}{0.58} & \textcolor{black}{4.5M}\\
CATH (exchange)
  & \textcolor{black}{\textbf{.301}} & \textcolor{black}{\textbf{.257}} & \textcolor{black}{\textbf{.271}}
  & \textcolor{black}{0.61} & \textcolor{black}{4.2M}\\
\bottomrule
\end{tabular*}
\end{table}

We use \textcolor{black}{frozen CLIP ViT-B/16 and SBERT all-mpnet-base-v2 encoders, SASRec, and Llama-3.1-8B-Instruct}, with ranking warm-up followed by fixed-executor PPO. Table~\ref{tab:config} lists training and execution settings. Online measurements use \textcolor{black}{4$\times$A100 80\,GB GPUs in BF16 with one query per worker}; routing, aggregation, and cache access enter the cost. Offline preparation is reported separately from serving.

\subsection{Ranking under Missing Content}
Table~\ref{tab:main} evaluates complete requests with \textcolor{black}{natural absence and no additional artificial masking}. HyperAgent4POI has the highest NDCG@20 among the prior baselines in this comparison. CATH improves NDCG@20 by \textcolor{black}{2.5pp on Gowalla-CA, 2.0pp on Foursquare-NYC, and 2.1pp on Foursquare-TKY}, corresponding to \textcolor{black}{8.4--9.1\% relative gains}. Recall@20 and MRR improve by \textcolor{black}{3.0--3.7pp and 1.6--1.9pp}. Independent scoring reaches \textcolor{black}{.282/.242/.255} NDCG@20, placing it between HyperAgent4POI and the full system. The following controls examine the group-selection and execution designs behind this comparison.

\begin{table}[!t]
\centering\footnotesize
\setlength{\tabcolsep}{2pt}
\caption{Complete-request quality and online cost on NYC (natural absence), with shared hardware, batching and concurrency. Tok.: processed tokens (K); Calls: tool/LLM calls.}
\label{tab:efficiency}
\setlength{\tabcolsep}{2pt}
\begin{tabular*}{\linewidth}{@{\extracolsep{\fill}}lcccccc@{}}
\toprule
Method & N@20 & Cost & p50 (ms) & p95 (ms) & Tok.\ (K) & Calls\\
\midrule
Frozen base
  & \textcolor{black}{.164} & \textcolor{black}{0.10}
  & \textcolor{black}{18} & \textcolor{black}{45}
  & \textcolor{black}{2.1} & \textcolor{black}{1}\\
HyperAgent4POI$^*$
  & \textcolor{black}{.237} & \textcolor{black}{0.22}
  & \textcolor{black}{8$^\dagger$} & \textcolor{black}{21$^\dagger$}
  & \textcolor{black}{31.4$^\ddagger$} & \textcolor{black}{0$^\dagger$}\\
MMA-Auto (adapt.)
  & \textcolor{black}{.226} & \textcolor{black}{0.46}
  & \textcolor{black}{124} & \textcolor{black}{318}
  & \textcolor{black}{18.7} & \textcolor{black}{4.2}\\
Independent
  & \textcolor{black}{.242} & \textcolor{black}{0.54}
  & \textcolor{black}{138} & \textcolor{black}{351}
  & \textcolor{black}{21.3} & \textcolor{black}{5.8}\\
Direct group
  & \textcolor{black}{.247} & \textcolor{black}{0.56}
  & \textcolor{black}{142} & \textcolor{black}{361}
  & \textcolor{black}{22.1} & \textcolor{black}{6.1}\\
Single controller
  & \textcolor{black}{.249} & \textcolor{black}{0.58}
  & \textcolor{black}{148} & \textcolor{black}{374}
  & \textcolor{black}{23.6} & \textcolor{black}{5.4}\\
CATH
  & \textcolor{black}{\textbf{.257}} & \textcolor{black}{0.61}
  & \textcolor{black}{156} & \textcolor{black}{398}
  & \textcolor{black}{24.8} & \textcolor{black}{6.7}\\
\bottomrule
\end{tabular*}
\par\raggedright\footnotesize
\textcolor{black}{$*$HyperAgent4POI excludes offline cache preparation ($\approx$6\,h); $^\dagger$cached serving without online LLM calls; $^\ddagger$offline preprocessing tokens.}
\end{table}

\begin{figure}[t]
\centering
\resizebox{\linewidth}{!}{
\begin{tikzpicture}[font=\scriptsize]
  \definecolor{robustheat}{RGB}{100,160,230}
  \definecolor{robustnumber}{RGB}{0,0,0}
  \def\cellw{1.10}
  \def\cellh{0.50}

  \node[anchor=west, font=\scriptsize\bfseries] at (0,0.85)
    {NDCG@20};
  \node[anchor=east, font=\tiny] at (4.40,0.85)
    {higher is better};
  \node[align=center] at (.55,.35) {Natural};
  \node[align=center] at (1.65,.35) {Text\\50\%};
  \node[align=center] at (2.75,.35) {Image\\50\%};
  \node[align=center] at (3.85,.35) {Joint\\50\%};

  \newcommand{\robustcell}[3]{%
    \pgfmathsetmacro{\heatamount}{10+80*(#3-.192)/(.301-.192)}
    \fill[robustheat!\heatamount, draw=white, line width=.7pt]
      ({(#2-1)*\cellw},{-#1*\cellh}) rectangle
      ({#2*\cellw},{-(#1+1)*\cellh});
    \ifnum#1=5
      \node[fill=none, inner sep=1.2pt, font=\scriptsize\bfseries,
        text=robustnumber] at
        ({(#2-.5)*\cellw},{-#1*\cellh-.5*\cellh}) {#3};
    \else
      \node[fill=none, inner sep=1.2pt, text=robustnumber] at
        ({(#2-.5)*\cellw},{-#1*\cellh-.5*\cellh}) {#3};
    \fi
  }
  \newcommand{\robustrow}[6]{%
    \ifnum#1=5
      \node[anchor=east, font=\scriptsize\bfseries] at
        (-.18,{-#1*\cellh-.5*\cellh}) {#2};
    \else
      \node[anchor=east] at
        (-.18,{-#1*\cellh-.5*\cellh}) {#2};
    \fi
    \robustcell{#1}{1}{#3}
    \robustcell{#1}{2}{#4}
    \robustcell{#1}{3}{#5}
    \robustcell{#1}{4}{#6}
  }
  \robustrow{0}{Frozen base}{.192}{.192}{.192}{.192}
  \robustrow{1}{HyperAgent4POI}{.276}{.261}{.268}{.248}
  \robustrow{2}{MMA-Auto (adapt.)}{.264}{.249}{.254}{.237}
  \robustrow{3}{Independent}{.282}{.266}{.273}{.254}
  \robustrow{4}{Direct group}{.281}{.265}{.271}{.253}
  \robustrow{5}{CATH}{.301}{.289}{.293}{.274}
  \draw[gray!40, thin] (0,0) rectangle (4.40,-3.00);
  \draw[black, semithick] (0,-2.50) rectangle (4.40,-3.00);

  \shade[left color=robustheat!10, right color=robustheat!90]
    (0,-3.33) rectangle (4.40,-3.47);
  \node[below, font=\tiny] at (0,-3.47) {.192};
  \node[below, font=\tiny] at (4.40,-3.47) {.301};
\end{tikzpicture}}
\caption{Ranking under missing content on Gowalla-CA. Cells show NDCG@20 under natural absence and 50\% masking of observed content, with shared requests and seeds. Color uses one scale across all cells. Frozen base relies solely on interaction IDs and is invariant to content masking.}
\label{fig:robustness}
\Description{Six methods are compared under natural absence, text masking, image masking, and joint masking. All 24 original scores are printed; CATH has the highest score in each condition, while the frozen base remains at .192.}
\end{figure}

Additional masking tests whether the ranking advantage persists when less content can be inspected. Because the frozen SASRec base retriever operates purely on sequential interaction IDs, its candidate ranking is invariant to multimodal masking (\textcolor{black}{.192} across all conditions on Gowalla-CA), serving as a content-agnostic baseline. On Gowalla-CA, gains over HyperAgent4POI range from \textcolor{black}{$+$2.5 to $+$2.8pp NDCG@20} across natural, text, image, and joint absence (Figure~\ref{fig:robustness}). Our score falls from \textcolor{black}{.301 to .274} under joint 50\% masking, while retaining \textcolor{black}{$+$2.6pp over HyperAgent4POI and $+$2.1pp over direct group selection}. The fixed candidate pools make these ranking differences independent of changes in retrieval coverage.

\subsection{Which Agents Should Analyze a Request Together?}
\label{sec:ablation}
W\begin{figure}[t]
\centering
\resizebox{\linewidth}{!}{
\begin{tikzpicture}[font=\scriptsize]
  \draw[->, thick, -latex] (0,0) -- (6.3,0) node[right, font=\scriptsize] {History stratum};
  \draw[->, thick, -latex] (0,0) -- (0,3.7) node[above, font=\scriptsize\bfseries] {NDCG@20};

  \foreach \y/\val in {0.0/0.24, 1.0/0.26, 2.0/0.28, 3.0/0.30} {
    \draw[gray!25, thin] (0,\y) -- (6.0,\y);
    \draw (0,\y) -- (-0.08,\y) node[left, font=\tiny] {\val};
  }
  \foreach \y in {0.5, 1.5, 2.5} {
    \draw[gray!15, very thin, dashed] (0,\y) -- (6.0,\y);
  }

  \definecolor{cgroupA}{RGB}{41,98,255}    
  \definecolor{cgroupB}{RGB}{230,81,0}     
  \definecolor{cgroupC}{RGB}{0,150,136}    
  \definecolor{cgroupD}{RGB}{123,31,162}   
  \definecolor{cours}{RGB}{190,20,30}      

  \def\barw{0.24}

  \fill[cgroupA!80, draw=cgroupA!90!black, thin] (0.53,0) rectangle (0.77,2.25);
  \draw[black, thick] (0.65,2.05) -- (0.65,2.45);
  \draw[black, thin] (0.60,2.05) -- (0.70,2.05);
  \draw[black, thin] (0.60,2.45) -- (0.70,2.45);

  \fill[cgroupB!80, draw=cgroupB!90!black, thin] (0.83,0) rectangle (1.07,1.15);
  \draw[black, thick] (0.95,0.95) -- (0.95,1.35);
  \draw[black, thin] (0.90,0.95) -- (1.00,0.95);
  \draw[black, thin] (0.90,1.35) -- (1.00,1.35);

  \fill[cgroupC!80, draw=cgroupC!90!black, thin] (1.13,0) rectangle (1.37,1.55);
  \draw[black, thick] (1.25,1.35) -- (1.25,1.75);
  \draw[black, thin] (1.20,1.35) -- (1.30,1.35);
  \draw[black, thin] (1.20,1.75) -- (1.30,1.75);

  \fill[cgroupD!80, draw=cgroupD!90!black, thin] (1.43,0) rectangle (1.67,1.90);
  \draw[black, thick] (1.55,1.70) -- (1.55,2.10);
  \draw[black, thin] (1.50,1.70) -- (1.60,1.70);
  \draw[black, thin] (1.50,2.10) -- (1.60,2.10);

  \node[below, font=\scriptsize\bfseries, align=center] at (1.1,-0.08) {Frequent\\Local};

  \fill[cgroupA!80, draw=cgroupA!90!black, thin] (2.43,0) rectangle (2.67,0.60);
  \draw[black, thick] (2.55,0.35) -- (2.55,0.85);
  \draw[black, thin] (2.50,0.35) -- (2.60,0.35);
  \draw[black, thin] (2.50,0.85) -- (2.60,0.85);

  \fill[cgroupB!80, draw=cgroupB!90!black, thin] (2.73,0) rectangle (2.97,2.10);
  \draw[black, thick] (2.85,1.85) -- (2.85,2.35);
  \draw[black, thin] (2.80,1.85) -- (2.90,1.85);
  \draw[black, thin] (2.80,2.35) -- (2.90,2.35);

  \fill[cgroupC!80, draw=cgroupC!90!black, thin] (3.03,0) rectangle (3.27,2.25);
  \draw[black, thick] (3.15,2.05) -- (3.15,2.45);
  \draw[black, thin] (3.10,2.05) -- (3.20,2.05);
  \draw[black, thin] (3.10,2.45) -- (3.20,2.45);

  \fill[cgroupD!80, draw=cgroupD!90!black, thin] (3.33,0) rectangle (3.57,1.05);
  \draw[black, thick] (3.45,0.80) -- (3.45,1.30);
  \draw[black, thin] (3.40,0.80) -- (3.50,0.80);
  \draw[black, thin] (3.40,1.30) -- (3.50,1.30);

  \node[below, font=\scriptsize\bfseries, align=center] at (3.0,-0.08) {Active\\Explorer};

  \fill[cgroupA!80, draw=cgroupA!90!black, thin] (4.33,0) rectangle (4.57,1.40);
  \draw[black, thick] (4.45,1.20) -- (4.45,1.60);
  \draw[black, thin] (4.40,1.20) -- (4.50,1.20);
  \draw[black, thin] (4.40,1.60) -- (4.50,1.60);

  \fill[cgroupB!80, draw=cgroupB!90!black, thin] (4.63,0) rectangle (4.87,1.95);
  \draw[black, thick] (4.75,1.75) -- (4.75,2.15);
  \draw[black, thin] (4.70,1.75) -- (4.80,1.75);
  \draw[black, thin] (4.70,2.15) -- (4.80,2.15);

  \fill[cgroupC!80, draw=cgroupC!90!black, thin] (4.93,0) rectangle (5.17,2.20);
  \draw[black, thick] (5.05,2.00) -- (5.05,2.40);
  \draw[black, thin] (5.00,2.00) -- (5.10,2.00);
  \draw[black, thin] (5.00,2.40) -- (5.10,2.40);

  \fill[cgroupD!80, draw=cgroupD!90!black, thin] (5.23,0) rectangle (5.47,1.20);
  \draw[black, thick] (5.35,1.00) -- (5.35,1.40);
  \draw[black, thin] (5.30,1.00) -- (5.40,1.00);
  \draw[black, thin] (5.30,1.40) -- (5.40,1.40);

  \node[below, font=\scriptsize\bfseries, align=center] at (4.9,-0.08) {Activity\\Streak};

  \draw[cours, thick, dashed] (0.3,2.65) -- (5.7,2.65) node[right, font=\tiny\bfseries, cours] {.293};
  \node[above, font=\tiny\bfseries, cours] at (3.0,2.65) {CATH (validation mean)};

  \draw[latex-latex, thick, blue!85!black] (0.65,2.50) to[bend left=25] node[above, font=\tiny\bfseries] {$-3.3$\,pp shift} (2.55,0.95);

  \begin{scope}[shift={(0.1, 3.25)}]
    \fill[cgroupA!80] (0.0,0) rectangle (0.22,0.14);
    \node[right, font=\tiny] at (0.22,0.07) {\tiny Traj,ST};

    \fill[cgroupB!80] (1.3,0) rectangle (1.52,0.14);
    \node[right, font=\tiny] at (1.52,0.07) {\tiny Pref,Text};

    \fill[cgroupC!80] (2.7,0) rectangle (2.92,0.14);
    \node[right, font=\tiny] at (2.92,0.07) {\tiny Pref,Text,Vis};

    \fill[cgroupD!80] (4.3,0) rectangle (4.52,0.14);
    \node[right, font=\tiny] at (4.52,0.07) {\tiny Collab,Traj};
  \end{scope}
\end{tikzpicture}}
\caption{Usefulness of agent groups across history strata on Gowalla-CA under joint 30\% masking. Bars: NDCG@20; errors: $\pm1$ seed std over 3 seeds. Traj/ST/Pref/Text/Vis/Collab denote trajectory, spatiotemporal, preference, text, visual and collaborative retrieval agents. Dashed line: validation mean (.293).}
\label{fig:groups}
\Description{The ordering of four agent groups varies across frequent-local, active-explorer, and activity-streak histories.}
\end{figure}
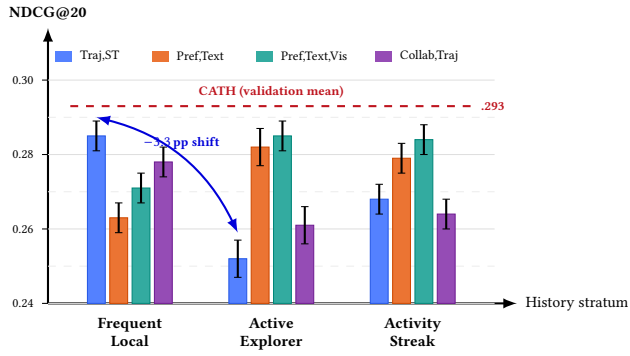

Before comparing selectors, we check their motivation: different histories may benefit from different agent groups.
Figure~\ref{fig:groups} shows that agent group usefulness varies across \textcolor{black}{frequent-local, active-explorer, and activity-streak} histories under controlled missingness, with \textcolor{black}{up to 3.3pp NDCG@20 variation for an individual group}, motivating history-dependent selection.
Table~\ref{tab:ablation}(a) compares selection rules with a common executor. Fixed-group selection uses a validation-selected set. Mask-only routing removes observed history, time, and location. Independent scoring reads full requests and masks but scores members unconditioned on the selected set. The direct selector evaluates legal sets with the fixed initiator via beam search.
Set-conditioned selection achieves the highest NDCG@20, exceeding independent scoring by \textcolor{black}{$+$1.4pp} and direct selection by \textcolor{black}{$+$1.2pp} on Gowalla-CA (NYC cost \textcolor{black}{0.61} vs.\ \textcolor{black}{0.54/0.56}), validating request-dependent group formation.

\subsection{Shared Results and Subsequent Operations}
Holding tasks, targets, and membership fixed, Table~\ref{tab:exchange}(b) examines how groups use evidence during execution. Both exchange and no-exchange allow multiple rounds; only exchange provides peer results to subsequent operations. The single controller uses identical operations and ranking interface over a flat history. Exchange outperforms local processing plus fusion by \textcolor{black}{$+$1.7pp NDCG@20} on Gowalla-CA and the single controller by \textcolor{black}{$+$0.9pp} (NYC costs \textcolor{black}{0.61 vs.\ 0.54 and 0.58}), confirming the benefit of shared intermediate results. \textcolor{black}{Retrieval targets or comparison sets change in 73\% of multi-round executions (365 out of 500 sampled 2-round episodes); in one representative trace, collaborative retrieval switches to a subcategory shortlist after receiving the text agent's POI-attribute summary,} demonstrating an actionable path from peer results to subsequent operations.

\subsection{Quality and Online Computation}
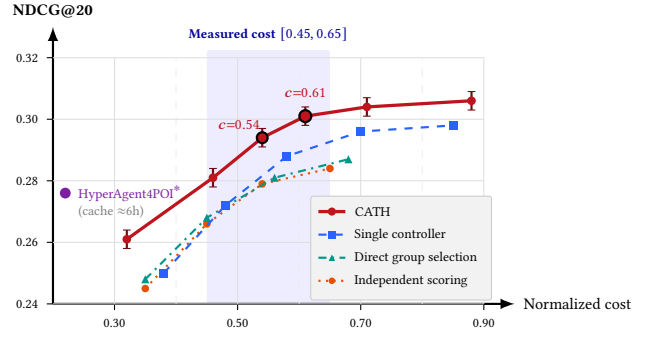
\begin{figure}[t]
\centering
\resizebox{\linewidth}{!}{
\begin{tikzpicture}[font=\scriptsize]

  \fill[blue!7, rounded corners=1pt] (2.0,0) rectangle (3.6,3.3);
  \node[above, font=\tiny\bfseries, text=blue!60!black] at (2.8,3.3) {Measured cost $[0.45, 0.65]$};

  \draw[->, thick, -latex] (0,0) -- (6.0,0) node[right, font=\scriptsize] {Normalized cost};
  \draw[->, thick, -latex] (0,0) -- (0,3.6) node[above, font=\scriptsize\bfseries] {NDCG@20};

  \foreach \x/\val in {0.8/0.30, 2.4/0.50, 4.0/0.70, 5.6/0.90} {
    \draw[gray!25, thin] (\x,0) -- (\x,3.2);
    \draw (\x,0) -- (\x,-0.08) node[below, font=\tiny] {\val};
  }
  \foreach \x in {1.6, 3.2, 4.8} {
    \draw[gray!15, very thin, dashed] (\x,0) -- (\x,3.2);
  }

  \foreach \y/\val in {0.0/0.24, 0.8/0.26, 1.6/0.28, 2.4/0.30, 3.2/0.32} {
    \draw[gray!25, thin] (0,\y) -- (5.6,\y);
    \draw (0,\y) -- (-0.08,\y) node[left, font=\tiny] {\val};
  }

  \definecolor{cours}{RGB}{190,20,30}       
  \definecolor{csingle}{RGB}{41,98,255}     
  \definecolor{cdirect}{RGB}{0,150,136}     
  \definecolor{cindep}{RGB}{230,81,0}       
  \definecolor{chyper}{RGB}{123,31,162}     

  \draw[cindep, thick, dotted] (1.20, 0.20) -- (2.00, 1.04) -- (2.72, 1.56) -- (3.60, 1.76);
  \foreach \px/\py in {1.20/0.20, 2.00/1.04, 2.72/1.56, 3.60/1.76} {
    \fill[cindep] (\px,\py) circle (1.4pt);
  }

  \draw[cdirect, thick, dashdotted] (1.20, 0.32) -- (2.00, 1.12) -- (2.88, 1.64) -- (3.84, 1.88);
  \foreach \px/\py in {1.20/0.32, 2.00/1.12, 2.88/1.64, 3.84/1.88} {
    \fill[cdirect] (\px,\py) ++(0,1.5pt) -- ++(-1.5pt,-3pt) -- ++(3pt,0) -- cycle;
  }

  \draw[csingle, thick, dashed] (1.44, 0.40) -- (2.24, 1.28) -- (3.04, 1.92) -- (4.00, 2.24) -- (5.20, 2.32);
  \foreach \px/\py in {1.44/0.40, 2.24/1.28, 3.04/1.92, 4.00/2.24, 5.20/2.32} {
    \node[fill=csingle, inner sep=0pt, minimum size=3.0pt] at (\px, \py) {};
  }

  \draw[cours, very thick] (0.96, 0.84) -- (2.08, 1.64) -- (2.72, 2.16) -- (3.28, 2.44) -- (4.08, 2.56) -- (5.44, 2.64);
  \foreach \px/\py in {0.96/0.84, 2.08/1.64, 4.08/2.56, 5.44/2.64} {
    \draw[cours!60!black, thick] (\px,\py-0.12) -- (\px,\py+0.12);
    \draw[cours!60!black, thin] (\px-0.05,\py-0.12) -- (\px+0.05,\py-0.12);
    \draw[cours!60!black, thin] (\px-0.05,\py+0.12) -- (\px+0.05,\py+0.12);
    \fill[cours] (\px,\py) circle (1.6pt);
  }
  \draw[cours!60!black, thick] (2.72,2.16-0.12) -- (2.72,2.16+0.12);
  \draw[cours!60!black, thin] (2.72-0.05,2.16-0.12) -- (2.72+0.05,2.16-0.12);
  \draw[cours!60!black, thin] (2.72-0.05,2.16+0.12) -- (2.72+0.05,2.16+0.12);
  \fill[cours, draw=black, thick] (2.72, 2.16) circle (2.0pt);
  \node[above left, font=\tiny\bfseries, cours] at (2.80, 2.16) {$c{=}0.54$};

  \draw[cours!60!black, thick] (3.28,2.44-0.12) -- (3.28,2.44+0.12);
  \draw[cours!60!black, thin] (3.28-0.05,2.44-0.12) -- (3.28+0.05,2.44-0.12);
  \draw[cours!60!black, thin] (3.28-0.05,2.44+0.12) -- (3.28+0.05,2.44+0.12);
  \fill[cours, draw=black, thick] (3.28, 2.44) circle (2.2pt);
  \node[above, font=\tiny\bfseries, cours] at (3.28, 2.58) {$c{=}0.61$};

  \fill[chyper] (0.16, 1.44) circle (2.0pt);
  \node[right, font=\tiny, chyper] at (0.22, 1.44) {HyperAgent4POI$^*$};
  \node[right, font=\tiny, gray] at (0.22, 1.20) {\tiny (cache $\approx$6h)};

  \begin{scope}[shift={(3.45, 0.15)}]
    \fill[white!95!black, draw=gray!40, rounded corners=1pt] (-0.1,-0.1) rectangle (2.25,1.25);
    \draw[cours, very thick] (0.0, 1.05) -- (0.35, 1.05) node[right, font=\tiny, text=black] {CATH};
    \fill[cours] (0.175, 1.05) circle (1.5pt);

    \draw[csingle, thick, dashed] (0.0, 0.75) -- (0.35, 0.75) node[right, font=\tiny, text=black] {Single controller};
    \node[fill=csingle, inner sep=0pt, minimum size=2.6pt] at (0.175, 0.75) {};

    \draw[cdirect, thick, dashdotted] (0.0, 0.45) -- (0.35, 0.45) node[right, font=\tiny, text=black] {Direct group selection};
    \fill[cdirect] (0.175, 0.45) ++(0,1.2pt) -- ++(-1.2pt,-2.4pt) -- ++(2.4pt,0) -- cycle;

    \draw[cindep, thick, dotted] (0.0, 0.15) -- (0.35, 0.15) node[right, font=\tiny, text=black] {Independent scoring};
    \fill[cindep] (0.175, 0.15) circle (1.2pt);
  \end{scope}
\end{tikzpicture}}
\caption{Complete-request quality versus measured computation on Gowalla-CA under natural missingness (matching Table~\ref{tab:main}). Cost is normalized to single-controller full pipeline cost. Error bars on CATH show $\pm1$ seed std over 3 seeds. Shading marks cost range $[0.45,0.65]$; operating points $c{=}0.54$ (.294 NDCG) and $c{=}0.61$ (.301 NDCG) reflect budget caps $B{=}0.54$ and $B{=}0.61$.}
\label{fig:cost}
\Description{NDCG@20 versus normalized computation for CATH, independent scoring, direct group selection, and a single controller; the cached baseline appears as a separate point.}
\end{figure}

We compare measured online computation across validation budget settings. Figure~\ref{fig:cost} plots quality--cost pairs on Gowalla-CA under natural missingness across budget caps $B \in \{0.30,\allowbreak 0.45,\allowbreak 0.54,\allowbreak 0.61,\allowbreak 0.72,\allowbreak 0.90\}$, normalized to single-controller cost. At normalized cost \textcolor{black}{0.54} ($B=0.54$), CATH exceeds independent scoring by \textcolor{black}{1.5pp NDCG@20}, with diminishing returns at higher budgets. Table~\ref{tab:efficiency} reports complete-request quality and online resources on NYC. CATH gains \textcolor{black}{0.8pp over the single controller and 1.5pp over independent scoring} (normalized cost \textcolor{black}{0.61 vs.\ 0.58 and 0.54}). Latency, processed tokens, and calls scale with this gain; HyperAgent4POI uses cached serving with separate offline preparation. The tradeoff guides budget selection under online constraints.
\section{Conclusion}
CATH uses task hypergraphs to structure multi-agent analysis around explicit tasks, connecting selected members, goals, targets, and shared candidate results. Across three benchmarks with incomplete multimodal content, the framework improves next-POI ranking under both natural and artificial missingness while characterizing computation tradeoffs.

\paragraph{Scope.}
Evaluation examines six agents on fixed candidate pools; retrieval misses remain outside the reranker. Untimestamped content is treated as static side information, and online cost reflects our hardware and caching configuration.

\clearpage
\balance
\bibliographystyle{ACM-Reference-Format}
\bibliography{reference}









\end{document}